\documentclass[twocolumn,prb,superscriptaddress]{revtex4-1}

\usepackage{amssymb}
\usepackage{amsmath}
\usepackage{graphicx}

\begin{document}

\title{Influence of spatial dispersion in metals on the optical response of deeply subwavelength slit arrays}

\author{Mathieu Dechaux}
\affiliation{Clermont Universit\'e, Universit\'e Blaise
  Pascal, Institut Pascal, BP 10448, F-63000 Clermont-Ferrand, France}
\affiliation{CNRS, UMR 6602, IP, F-63171 Aubi\`ere, France} 
\author{Paul-Henri Tichit}
\affiliation{Clermont Universit\'e, Universit\'e Blaise
  Pascal, Institut Pascal, BP 10448, F-63000 Clermont-Ferrand, France}
\affiliation{CNRS, UMR 6602, IP, F-63171 Aubi\`ere, France} 
\author{Cristian Cirac\`i}
\affiliation{Instituto Italiano di Tecnologia (IIT), Center for Biomolecular Nanotechnologies, Via Barsanti, I-73010 Arnesano, Italy}
\author{Jessica Benedicto}
\affiliation{Clermont Universit\'e, Universit\'e Blaise
  Pascal, Institut Pascal, BP 10448, F-63000 Clermont-Ferrand, France}
\affiliation{CNRS, UMR 6602, IP, F-63171 Aubi\`ere, France}
\author{R\'emi Poll\`es}
\affiliation{Clermont Universit\'e, Universit\'e Blaise
  Pascal, Institut Pascal, BP 10448, F-63000 Clermont-Ferrand, France}
\affiliation{CNRS, UMR 6602, IP, F-63171 Aubi\`ere, France}  
\author{Emmanuel Centeno}
\affiliation{Clermont Universit\'e, Universit\'e Blaise
  Pascal, Institut Pascal, BP 10448, F-63000 Clermont-Ferrand, France}
\affiliation{CNRS, UMR 6602, IP, F-63171 Aubi\`ere, France} 
\author{David R. Smith}
\affiliation{Center for Metamaterials and
  Integrated Plasmonics, Duke University, Durham, North Carolina
  27708, USA} 
\author{Antoine Moreau}
\affiliation{Clermont Universit\'e, Universit\'e Blaise
  Pascal, Institut Pascal, BP 10448, F-63000 Clermont-Ferrand, France}
\affiliation{CNRS, UMR 6602, IP, F-63171 Aubi\`ere, France} 
\affiliation{Center for Metamaterials and
  Integrated Plasmonics, Duke University, Durham, North Carolina
  27708, USA} 

\begin{abstract}
In the framework of the hydrodynamic model describing the response of electrons in a metal, 
we show that arrays of very narrow and shallow metallic slits have an optical response that
is influenced by the spatial dispersion in metals arising from the repulsive interaction
between electrons. As a simple Fabry-Perot model is not accurate enough to describe the structure's behavior, we 
propose to consider the slits as generalized cavities with two modes, one being propagative and
the other evanescent. This very general model allows to conclude that the impact of spatial dispersion
on the propagative mode is the key factor explaining why the whole structure is sensitive to
spatial dispersion. As the fabrication of such structures with relatively large gaps compared to previous experiments
is within our reach, this work paves the way for future much needed experiments on nonlocality. 
\end{abstract}

\maketitle

\section{Introduction}

Drude's  model\cite{drude00}   has  proven  unbelievably   accurate  throughout  the
twentieth century to describe  the optical response of metals, despite
extensive studies  in the seventies  and eighties to find  its limits\cite{boardman82,forstmann86}.
Recent experiments have however shown that the behavior of resonances
in  sub-nanometer-sized gaps can  not be  explained with  Drude's model
alone\cite{ciraci12,ciraci14}, making it necessary to  take into account the repulsion between
electrons inside the  metal in the framework of  a hydrodynamic model\cite{crouseilles08,moreau13,ciraci13}.   In that  case,  the  metallic  response is  spatially
dispersive and  since it  cannot be reduced  to a  single permittivity
depending  on   the  frequency   alone,  it  is   often  said   to  be
non-local. Further experiments would however be welcome because of the
sub-nanometer dimensions of the considered gaps\cite{teperik13,hajisalem14}.

In the present  work, we explain why the  non-local response of metals
can be  expected to  have an impact  on deeply  subwavelength metallic
gratings, a  structure that has  been extensively studied in  the past
decade for its extraordinary transmission\cite{porto99,huang10} and absorption\cite{leperchec08,pardo11,polyakov11,polyakov12}. We underline
that these effect will be clear for grooves that are as large as a few
nanometers and  that the fabrication  of such still very  narrow slits
seems totally within our reach\cite{polyakov11,polyakov12,chen14}.

\begin{figure}[htb]
\centerline{\includegraphics[width=7cm]{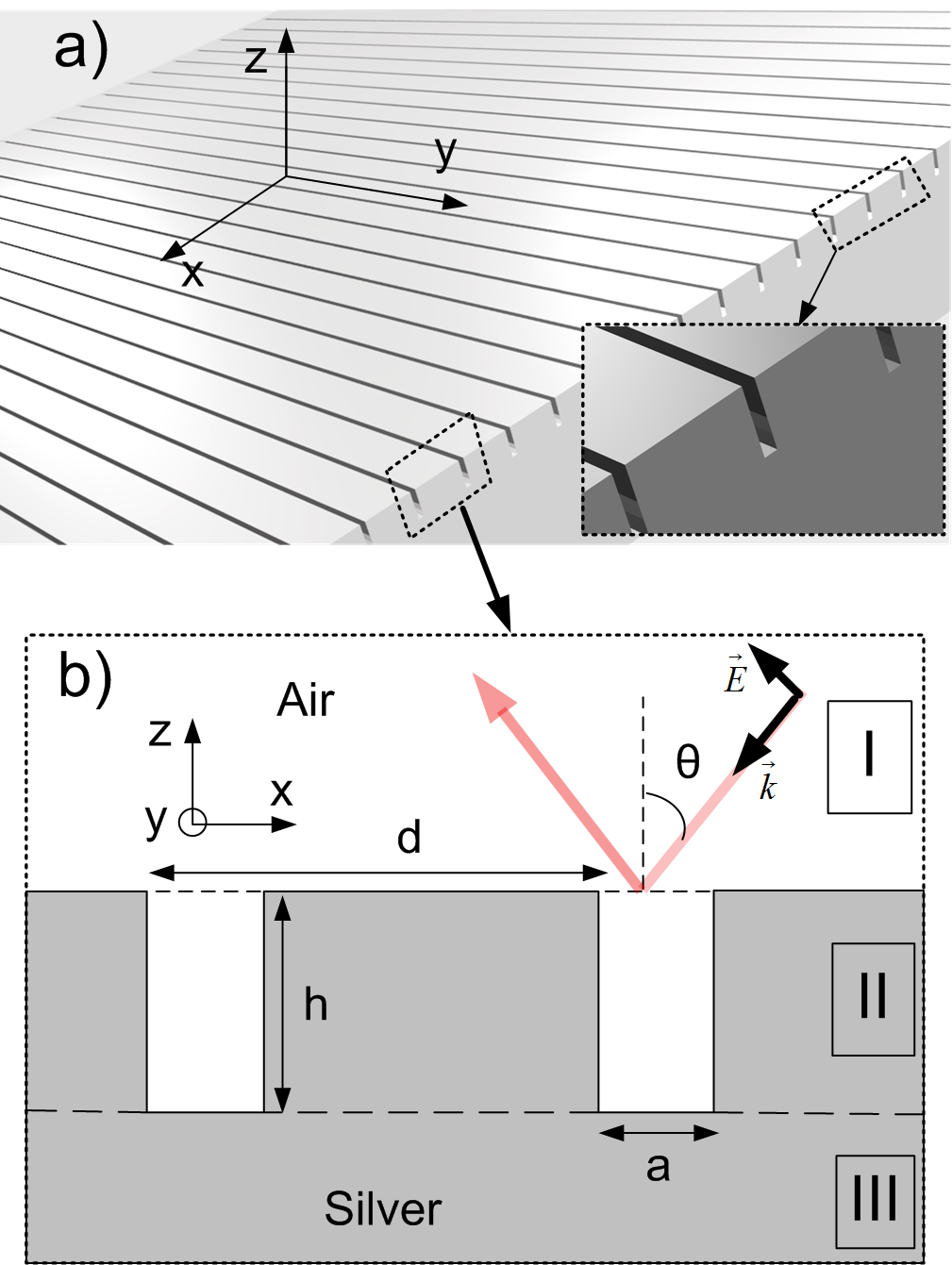}}
\caption{(Color online) Representation of the grooves carved in metal,
  illuminated from above by  a TM-polarized plane wave. We distinguish
  three  different layers,  layer II  being the  layer  containing the
  slits.\label{fig:schema}}
\end{figure}

In  the first  part of  this article,  we will  focus on  the physical
analysis of  the absorption by  the metallic slits array  when spatial
dispersion   is  neglected.  It   is  now   well  accepted   that  the
extraordinary  optical  transmission of  slit  arrays  is  due to  the
excitation of cavity resonances inside the slits\cite{porto99,lalanne00}, even if non-resonant
mechanism allow for a high transmission for very thin structures
\cite{subramania11}. The only guided mode
propagating in the slits in p-polarization has actually no cut-off and
can propagate whatever  the slit width, with a  wavevector $k_z$ close
to $k_0$, the  wavevector in vacuum in most cases. This  is why, as is
usual for cavities,  the thickness of the grating  is roughly half the
wavelength  in  vacuum for the fundamental resonance\cite{marquier05}, except for exotic cases\cite{moreau07}. 
The  same mechanism  explains  the absorption by subwavelength grooves,  the difference being that, since
the cavity is  now closed on one end  (see Fig. \ref{fig:schema}), the
depth of the grooves is only a quarter of the wavelength in vacuum\cite{pardo11}. In
all cases,  an extraordinarily strong funneling effect  explains the way
the energy is literally sucked up into the slits\cite{subramania11,pardo11}.

As long as the metal can be considered as almost perfect (in the IR\cite{huang10,pardo11} or
THz\cite{fan12} or even microwave\cite{akozbek12} range) or when the slit is larger than 50 nm in the optical range,
the  above  physical picture  is  fully  accurate. However,  below  that  50-nm
threshold for the slit width in  the optical range, the guided mode is
more localized in the metal than in the slits because the skin depth $\delta$ is
typically around 25 nm for noble  metals. The mode  then experiences what
can be called a plasmonic drag:  as the width of the gap decreases, it
becomes slower  and slower,  its group velocity  goes to zero  and its
wavevector  $k_z$  diverges\cite{bozhevolnyi07}.  Its  effective  wavelength,  defined  as
$\lambda_{eff}=2\pi/k_z$  thus becomes  very  small. The depth of  the
cavity being  actually proportional to this  effective wavelength\cite{yan12}, the
thickness of  the grating  can be made  extremely small and  the slits
will  still  constitute  a  resonator\cite{leperchec08}.  This  is  the  regime  we  are
interested in,  because it is  one of the  rare structures in  which a
plasmonic guided mode with a  very high wavevector can be excited\cite{moreau13}.  We
show however  that the  physics of these  resonators is  slightly more
complicated that previously thought\cite{leperchec08}: the slits are so shallow that the
guided mode is  not the only channel for light to  reach the bottom of
the  structure. A  one-mode  model\cite{porto99,lalanne00} is  thus  not sufficient  to
describe  the  cavity accurately.  We  give  here  a generalized  Fabry-Perot
formula to better predict the behavior of the resonances.

In  a second  part, we  take into  account spatial  dispersion  in the
framework of the hydrodynamic model with realistic parameters\cite{moreau13,benedicto15}, and show that the response of the
structure   is  influenced   by  nonlocality.   Moreover,   using  the
generalized Fabry-Perot formula, we show that the influence of nonlocality on
the  wavevector of  the guided  mode explains  almost totally  why the
structure is so sensitive  to spatial dispersion in metals, completing
the physical picture.

\section{Generalized cavity model}

The structure considered here\cite{leperchec08,polyakov11,polyakov12} is presented in  Figure \ref{fig:schema}. It is  an infinite
array of  deeply subwavelength grooves of width  $a$ ranging 2 to  5 nm and depth $h$, from 15 to 30 nm typically.
The period $d$ is ranging from 30 to 50 nm in the present study. Since the structure is periodic with
a period of the order of twice the skin depth, the slits must be considered as coupled, and the modes
propagating in the slits can be found for normal incidence by solving the classical dispersion relation for
metallo-dielectric structures\cite{leperchec08}
\begin{multline}\label{eq:disp}
\cosh(\kappa\,                    a)                    \cosh(\kappa_t
(d-a))\\                     +\frac{1}{2}\left[\frac{\kappa_t}{\epsilon
    \kappa}-\frac{\epsilon  \kappa}{\kappa_t}\right]  \sinh(\kappa  a)
\sinh\left(\kappa_t(d-a)\right)-1 = 0,
\end{multline}
where  $\kappa_t^2=k_n^2-\epsilon\,k_0^2$  and
$\kappa^2=k_n^2-k_0^2$, $\epsilon$ being the permittivity of the metal
and $k_0=2\pi/\lambda$.

The different modes are indexed by $n$ and characterized by a propagation constant 
$k_z$ along the $z$ axis, and a magnetic field profile $H_y^n(x)$. The magnetic field
in layer II can thus be written as a sum of modes propagating upward or downward
\begin{equation}
H_y(x,z)  =  \sum_{n}  H_y^n(x) \left[A_n  \,e^{-i\,k_nz}+B_n\,e^{+i\, k_nz}\right]e^{-i\omega\,t}.\label{e:form}
\end{equation}

\begin{figure}[htb]
\centerline{\includegraphics[width=8cm]{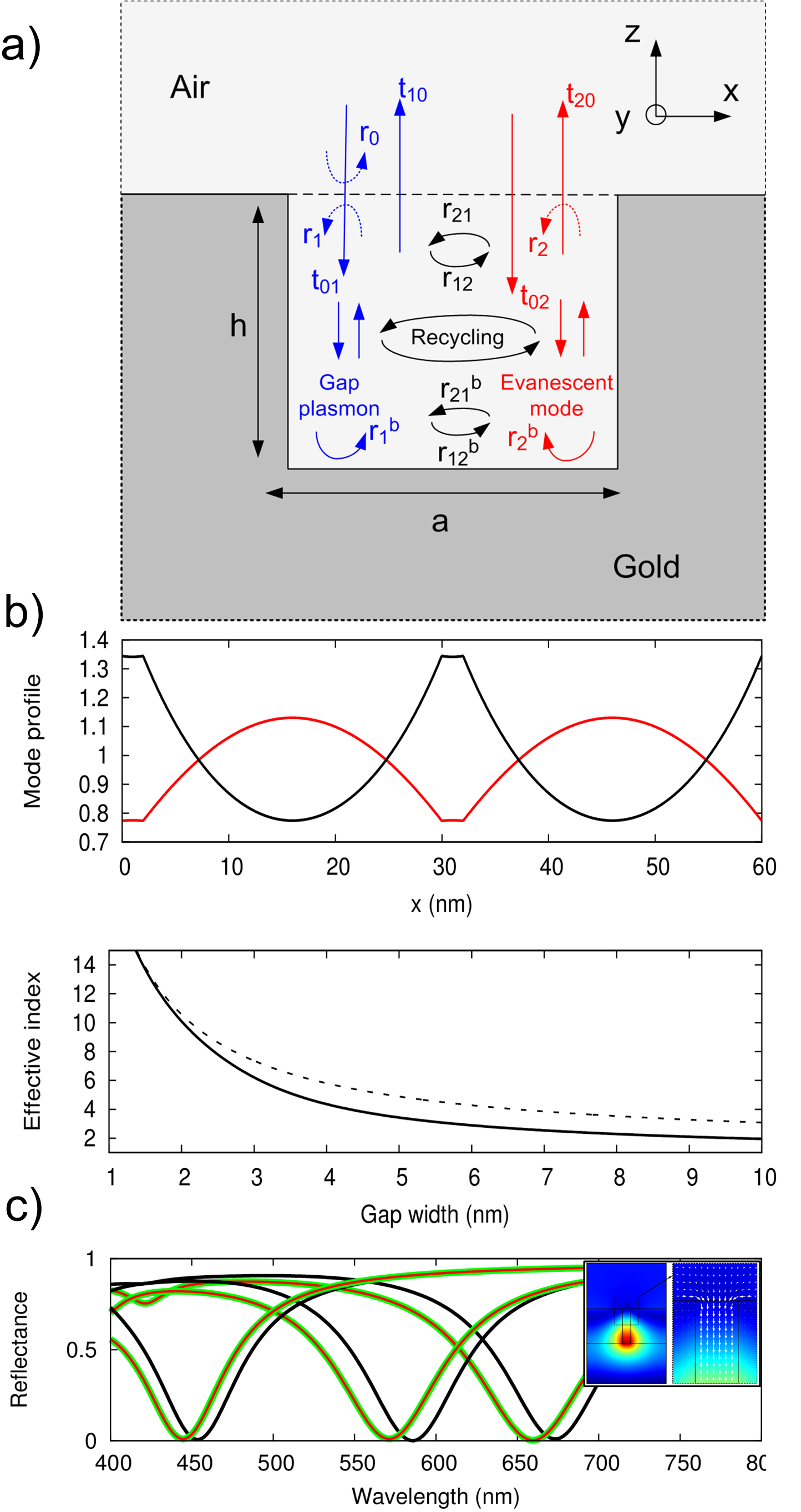}}
\caption{(Color online) (a) Representation of the generalized cavity model with the different reflection coefficients. (b) Top: Profile of the propagating mode (black curve) and the first evanescent mode(red curve). Bottom: Diverging effective index of the propagating mode as a function of the gap width (solid line) and the mode guided in a single isolated slit, the gap-plasmon polariton (dotted line). (c) Reflection coefficient of the whole structure, considering only the fundamental Bloch mode (black curve), a 2 modes model (red curve) and a complete simulation (green curve) for different groove depth (left: 7nm; center: 3 nm; right: 2 nm). Inset: simulation with Comsol multiphysics of the magnetic field and the Poynting vector at the entrance of the slits.\label{f:2}}
\end{figure}

As expected, there  is only one  propagating Bloch  mode (presenting  a propagation
constant $k_1$ with  a dominant real part). All the other modes are evanescent and
thus attenuated in the $z$ direction, with essentially imaginary propagation constants $k_n$. 
The propagative mode is reflected back and forth in the
slits, thus producing the resonance. The actual reflection coefficients ($r_1$ for the interface between the grating and air,
$r_1^b$ for the bottom of the grooves) can be computed using RCWA\cite{lalanne96,granet96}, as is quite common for metallic gratings\cite{lalanne00,moreau07}.
We underline that the computation of the reflection coefficients requires the computation of many evanescent modes. One could
expect, from the vast literature on the subject, that the reflection coefficient of the whole structure can simply be written
using a Fabry-Perot formula:
\begin{equation}
r=r_0+\frac{r_1\,t_{01}\,t_{10}\,e^{2ik_1    h}}{1-r_1\;r^b_1\,e^{2ik_1 \,h}},\label{e:1mode}
\end{equation}
where $t_{01}$ is the transmission coefficient from the incoming plane wave to the propagating mode in the slits and
$t_{10}$ the transmission coefficient from the mode in the slits to the outgoing plane wave, that are computed using
RCWA. While such an approach has proved extremely accurate in the past for the Extraordinary Optical Transmission (EOT) \cite{porto99,lalanne00,moreno01,moreau07},
here, quite unexpectedly, it fails to predict the position of the resonance given by local RCWA full simulations (see Fig. \ref{f:2}).

The assumption underlying \eqref{e:1mode} is that only the propagative mode is able to reach the bottom
of the grooves. However, because the depth of the grating is smaller than the skin depth, not all the evanescent modes are attenuated enough to be neglected.
To be more precise, one mode in particular, although it is not propagative in the $z$ direction, presents an attenuation constant that is so low that it is still significantly strong at the bottom of the slits. 

 The profile of this mode is shown Fig. \ref{f:2} and it is relatively flat. This mode
is thus very close to the attenuated wave that is excited in the metal when a plane wave is
reflected by a metallic screen. Moreover, each reflection of the propagating mode at the top or at the bottom
of the grooves generates this mode too, so that it fully participates to the resonance. We propose here
a very general two-modes approach to model the reflection coefficient of the structure.

In this framework, we call $A_1$ and $B_1$ (resp. $A_2$  and $B_2$) are the
amplitude  of  the  propagating  (resp. evanescent)  mode traveling (resp. attenuating)
downwards or upwards (see Fig. \ref{f:2}). The reflection coefficient of the whole structure can then  be written as the  result of
the reflection on the metallic  plane plus  the light  that comes  from the two modes inside the grating layer
\begin{equation}
r = r_0+ t_{10} B_1 +t_{20}\,B_2.
\end{equation}
The downward amplitudes can be written as a result of the direct excitation by the incoming plane wave, in addition to the reflection of the modes inside the slits
\[
\left\{
\begin{aligned}
A_1 = t_{01}+r_1\,B_1 + r_{21}\, B_2\\
A_2 =t_{02} + r_2\,B_2+r_{12}\,B_2
\end{aligned}
\right.
\]
and the equivalent is obtained at the bottom of the grooves for the upward amplitudes
\[
\left\{
\begin{aligned}
B_1\,e^{-ik_1h} = r_1^b\,A_1 \,e^{ik_1h} + r_{21}^b\, B_2\,e^{ik_2h}\\
B_2 \,e^{-ik_2h}= r_2^b\,A_1 \,e^{ik_1h}+r_{12}^b\,A_2\,e^{ik_2h}.
\end{aligned}
\right.
\]
This forces us to introduce all the above new reflection and coupling coefficients, that can be computed using RCWA. This approach is inspired by mode recycling in photonic crystal cavities\cite{sauvan05}, generalized here to (i) a non symmetrical plasmonic cavity and (ii) the case where the evanescent mode can be directly excited by the incident wave. 

We first use the relation above to eliminate $B_2$ from the whole system, yielding
\[  
\left\{
\begin{aligned}
 r&=r_0+t_{10}B_1+t_{20}(r_2^bA_2 e^{2ik_2h}+r_{12}^bA_1e^{i(k_1+k_2)h})\\
 A_1&=t_{01}+r_1\,B_1+r_{21}(r_2^b\,A_2 e^{2ik_2h}+r_{12}^bA_1e^{i(k_1+k_2)h})\\
 A_2&=t_{02}+r_1(r_2^bA_2 e^{2ik_2h}+r_{12}^bA_1e^{i(k_1+k_2)h})+r_{12}B_1\\
  B_1&=r_1^bA_1 e^{2ik_1h}+r_{21}^bA_2e^{i(k_1+k_2)h}
  \end{aligned}
\right.
\]

We finally get from these equations an expression for $A_2$
\[
A_2 = t_{02}'+c\,A_{1}+r_{12}'B_{1}
\]
where
\[\begin{aligned}
t_{02}'&=\frac{t_{02}}{1-r_2r_2^be^{2ik_2h}}\\
c&=\frac{r_2r_{12}^be^{i(k_2+\gamma_1)h}}{1-r_2 r_2^be^{2ik_2h}}\\
r_{12}'&=\frac{r_{12}}{1-r_2r_2^be^{2ik_2h}}.
\end{aligned}\]

We can thus replace $A_2$ in the above equations, to yield
\[  
\left\{
\begin{aligned}
 r&=r_{0}'+ t_{10}'B_{1}+\alpha A_{1}\\
 A_1&=t_{01}+r_1B_1+r_{21}(r_2^b (t_{02}'+cA_{1}\\
 &+r_{12}'B_{1}) e^{2ik_2h}+r_{12}^bA_1e^{i(k_1+k_2)h})\\
  B_1&=r_1^bA_1 e^{2ik_1h}+r_{21}^b(t_{02}'+cA_{1}+r_{12}'B_{1})e^{i(k_1+k_2)h}\\
  \end{aligned}
\right.
\]
where
\[\begin{aligned}
 r_{0}'&=r_0+t_{20}r_2^bt_{02}'e^{2ik_2h}\\
 t_{10}'&=t_{10}+t_{20}r_2^br_{12}'e^{2ik_2h}\\
 \alpha&=t_{20}r_{12}^be^{i(k_1+k_2)h}+t_{20}r_2^bc e^{2ik_2h}
\end{aligned}\]

And finally, the system reduces to 
\[  
\left\{
\begin{aligned}
 r&=r_{0}'+ t_{10}'B_{1}+\alpha A_{1}\\
 A_1&=r_1'B_1+ t_{01}'\\
 B_1&=r_1^{b'}A_1e^{2ik_1h}+ t_{02}^{''}\\
 \end{aligned}
\right.
\]
where 
\[\begin{aligned}
r_1^{'}&=\frac{r_2^br_{21}r_{12}'e^{2ik_2h}}{1-r_{21}r_{12}^be^{i(k_1+k_2)h}}\\
t_{01}'&=t_{01}+t_{02}'r_{21}r_2^be^{2ik_2h}\\
r_1^{b'}&=\frac{r_1^br_{21}^bce^{i(k_2-\gamma_1)h}}{1-r_{21}^br_{12}^{'-}e^{i(k_2-k_1)h}}\\
t_{02}^{''}&=\frac{t_{02}'r_{21}^be^{i(k_2+\gamma_1)h}}{1-r_{21}^br_{12}^{'-}e^{i(k_2+k_1)h}}\\
\end{aligned}\]

It is not obvious yet that these equations can be used to yield a generalized Fabry-Perot formula, because two terms have appeared that do not exist in the classical one-mode model. More precisely, to retrieve the exact same equations, $t_{02}^{''}$ and $\alpha$ would have to vanish. This is usually the case in photonic crystals - but physically here, the evanescent mode is directly and efficiently excited by the incoming wave so that $t_{02}^{''}$ is not negligible. 

We now replace $A_1$ by its expression in the reflexion coefficient's formula, to yield
\[
\left\{
\begin{aligned}
  r&=r_{0}^{''}+ t_{10}^{''} B_{1}\\
  B_1&=r_1^{b'}e^{2ik_1h} \,A_1e^{2ik_1h}+ t_{02}^{''}\\
  A_1&=r_1'B_1+ t_{01}'\\
  \end{aligned}
\right.
\]
where new effective coefficients are introduced: 
\begin{equation}
r_{0}^{''}=r_{0}'+\alpha  t_{01}'
\end{equation}
\begin{equation}
t_{10}^{''}=t_{10}'+\alpha  r_1'
\end{equation}

Finally, $A_1$ can be written 
\begin{equation}
A_1=\frac{t_{01}'+ t_{02}^{''}r_1'}{1-r_1'r_1^{b'}e^{2ik_1h}}
\end{equation}
and the reflection coefficient, by eliminating $B_1$ gives
\begin{equation}
r=r_{0}^{''}+ t_{10}^{''} r_1^{b'}A_1e^{2ik_1h}+ t_{10}^{''}t_{02}^{''}
\end{equation}
leading to the desired result
\begin{equation}
r=r_{0}^{''}+ t_{10}^{''}t_{02}^{''}+ \frac{r_1^{b'}e^{2ik_1h}t_{10}^{''}(t_{10}'+t_{02}^{''}r_1')}{1-r_1'r_1^{b'}e^{2ik_1h}}.
\end{equation}

If we call
\begin{equation}
r_{eff}=r_{0}^{''}+ t_{10}^{''}t_{02}^{''}
\end{equation}
\begin{equation}
t_{eff}=t_{10}^{''}(t_{10}'+t_{02}^{''}r_1')
\end{equation}

Then the reflection coefficient can be put under a form similar to the Fabry-Perot formula
\begin{equation}
r=r_{eff}+ \frac{t_{eff}r_1^{b'}e^{2ik_1h}}{1-r_1'r_1^{b'}e^{2ik_1h}}. \label{e:generalized}
\end{equation}

This formula can be considered as a generalized Fabry-Perot formula. The agreement between this formula and a full RCWA simulation (see Fig. \ref{f:2} (c)), that can be considered as a multi-mode model, is excellent. The two modes are thus the only ones that are responsible for the resonance. More precisely, the propagating mode is responsible for the resonance and the evanescent mode is responsible for a shift of this resonance compared to the one-mode model.

\begin{figure}[!h]
\begin{center}
\includegraphics[width=8cm]{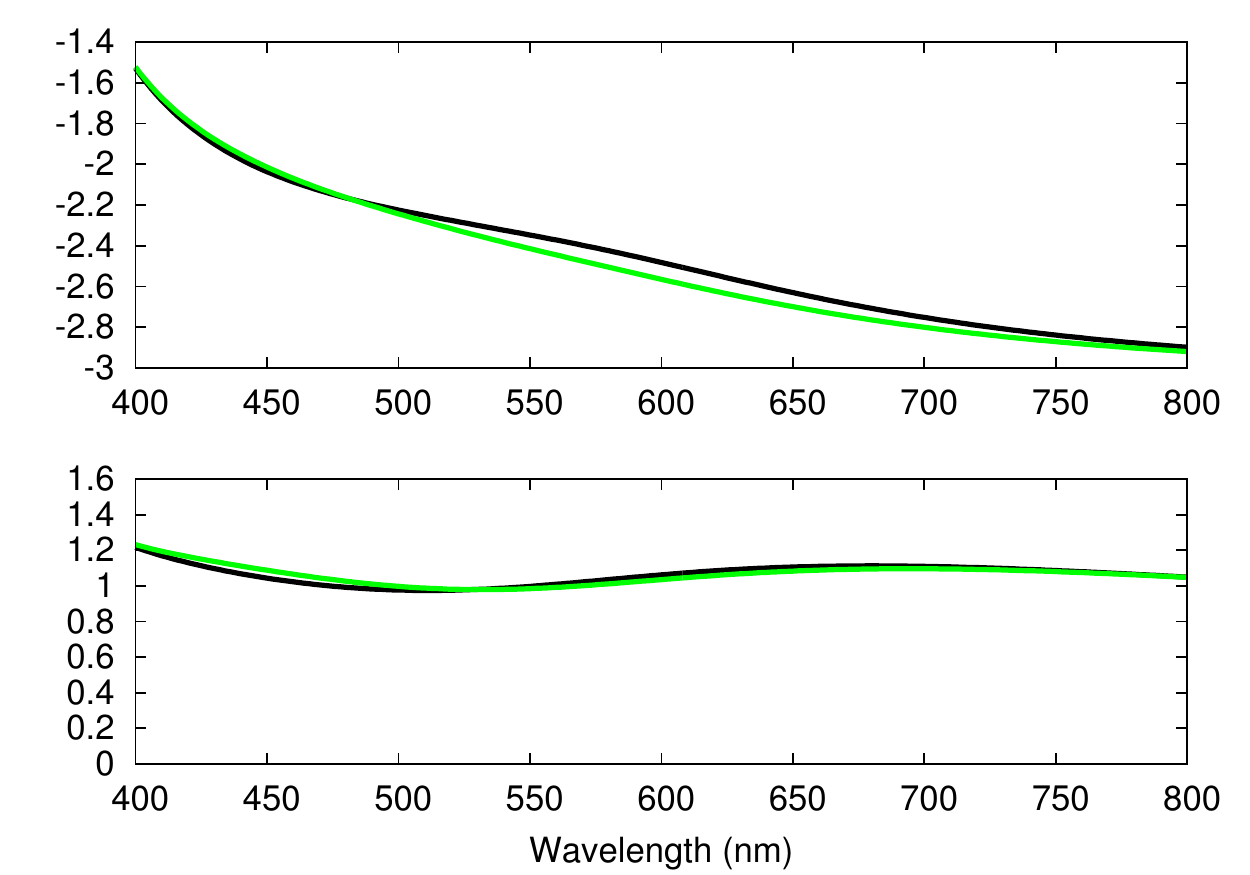}
\caption{(Color online) {Phase of the reflection coefficients.} Top: Phase of $r_1$ and $r_1'$ (black and green curves respectively). Bottom:  Phase of $r_1^b$ and $r_1^{b'}$ (black and green lines resp.). \label{f:phase}}
\end{center}
\end{figure}

The effective reflection coefficients are easy to compute, once the real coefficients are extracted from the scattering matrices of the interfaces between the different space regions\cite{moreau07}. The resonance condition now reads 
\begin{equation}
\arg(r_1') + \arg(r_1^{b'}) + 2\Re (k_1) h = 2m\pi
\end{equation}
where $m$ is a relative integer. Fig. \ref{f:phase} shows the phase of the effective reflection coefficients, compared to the phase of the real reflection coefficients. There is essentially a shift of the phase of $r_1'$ compared to the one of $r_1$. This totally explains why, compared to the one-mode model prediction, the resonance is shifted since the shift in the phase has a direct impact on the resonance condition. The fact that the slope of the phase is not changed means that there is almost no impact on the quality factor of the resonance, so that the only impact of the second mode on the resonance is to shift it without changing its width.

\section{Impact of spatial dispersion}

Deeply subwavelength gratings are very interesting because the absorbing resonance is the sign that a guided mode with a very high wavevector has been efficiently excited in the structure. Such modes are likely to be influenced by the repulsion between electrons\cite{moreau13} because their effective wavelength is so small that is approaches the mean free path of electrons in the metal\cite{chapuis08}, which is the relevant scale for nonlocality. That is the reason why the spatial dispersion is expected to have an impact on the slit array's response.

We use here the hydrodynamic model\cite{ciraci13,moreau13} to take the intrinsic nonlocality of the metallic response into account. The currents ${\bf J}$ corresponding to the movement of the free electrons trapped in the metal can be taken into account as an effective polarization ${\bf P}_f$ defined by ${\dot{\bf P}_f}={\bf J}$. The electron gas can be considered as a fluid\cite{crouseilles08}, leading to the following linearized equation
\begin{equation}\label{eq:hydro}
-\beta^2\nabla\left(\nabla.{\bf P}_f\right)+\ddot{\bf P}_f+\gamma\dot{\bf P}_f=\epsilon_0\, \omega_p^2\,{\bf E}
\end{equation}
where $\gamma$ is the damping factor, $\omega_p$ the plasma frequency and $\beta\simeq1.35.10^6$m/s. All these parameters (except $\beta$), as well as the dispersive susceptibility $\chi_b$ due to interband transitions are taken from careful fits of the available experimental data using a Drude and Brendel-Bormann model\cite{rakic98}. This allows us to clearly distinguish between the response of the jellium and the response of the background, that we assume is local. Maxwell's equations then reduce to\cite{moreau13}
\[  
\left\{
\begin{aligned}
 \nabla\times{\bf E}=i\omega\mu_0{\bf H}\\
 \nabla\times{\bf H}=-i\omega\epsilon_0(1+\chi_b){\bf E}+{\bf P}_f
\end{aligned}
\right.
\]
where, thanks to \eqref{eq:hydro} the polarization can be written as
\begin{equation}
{\bf P}_f=\frac{\epsilon_0.\omega^2_p}{\omega^2+i\gamma\omega}\left({\bf E}-(1+\chi_b)\frac{\beta^2}{\omega^2_p}\nabla\left(\nabla.{\bf E}\right)\right).
\end{equation}

The resolution of such equations in a multilayered structure requires the introduction of an additional boundary condition (ABC). The most obvious ABC is in that case to consider that no electrons are allowed to get out of the metal\cite{yan12}, which means ${\bf J}.{\bf n} = {\bf P}_f.{\bf n} =0$, where ${\bf n}$ is the normal to the interface - and it turns out to be one of the most conservative ABC, so that nonlocal effects are not likely to be overestimated\cite{moreau13}.

\begin{figure}[htb]
\centerline{\includegraphics[width=8cm]{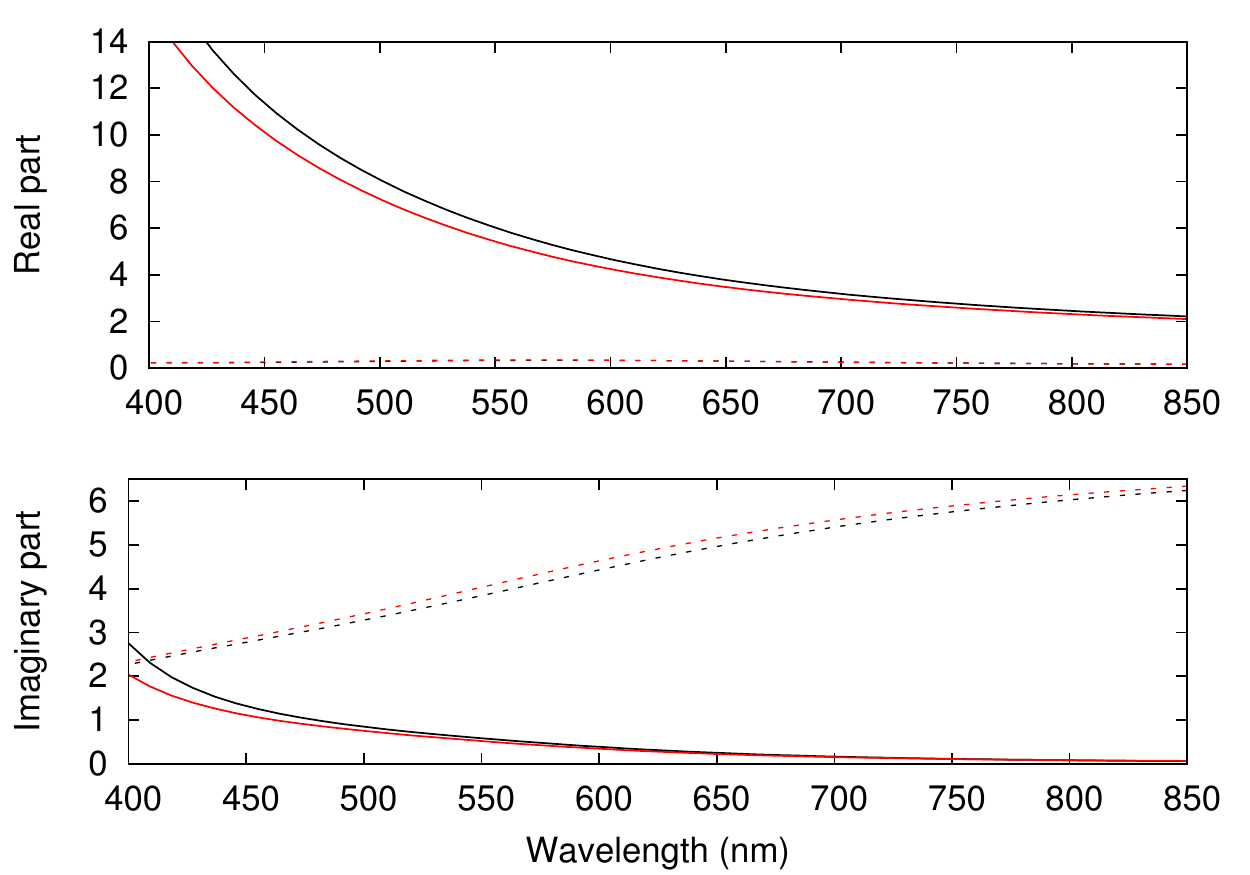}}
\caption{\label{fig:prop} (Color Online) Impact of nonlocality on the effective index (phase index) of the propagating mode (solid line) and on the evanescent mode (dashed line). The real (top) and imaginary (bottom) parts of the propagation constant are shown for a purely local approach (in black) and in the framework of the hydrodynamic model (red).}
\end{figure}

In this framework the dispersion relation giving the propagation constant of the modes is modified\cite{benedicto15} and becomes
\begin{multline}\label{eq:dispnl}
1-\left[\frac{\epsilon\Omega\sinh(\kappa_t e)}{k_z\sinh(\kappa_l e)}\right]=\cosh(\kappa a)\cosh(\kappa_t e)\\+\frac{1}{2}\left[\frac{\kappa_t}{\epsilon \kappa a}-\frac{\epsilon \kappa a}{1-\kappa_t}+\Omega^2\left(\frac{\epsilon}{\beta \kappa_t e}\right)\right]\sinh(\kappa a)\sinh(\kappa_t e)\\+\frac{\Omega}{\sinh(\kappa_l e)}\left[\frac{\sinh(\kappa a)}{\beta}\left(1-\cosh(\kappa_t e)\cosh(\kappa_l e)-\right.\right.\\\left.\left. \frac{\epsilon\sinh(\kappa_t e)}{\kappa_t}\cosh(\kappa a)\cosh(\kappa_t e)\right)\right]
\end{multline}
where  $e=d-a$ , $\Omega=\frac{k_z^2}{\kappa_l}(\frac{1}{\epsilon}-\frac{1}{1+\chi_b})$  , $\kappa_l^2=k_z^{2}+\Big(\frac{\omega_p^{2}}{\beta^{2}}\Big)\Big(\frac{1}{\chi_f}+\frac{1}{1+\chi_b}\Big)$ , $\kappa_t^{2}=k_z^{2}-\epsilon\,k_0^{2}$ and $\kappa^{2}=k_z^{2}-k_0$. This allows to consider the impact of nonlocality on the propagation constants of the two modes that are involved in the resonance of the structure - both the propagating and the least evanescent of the remaining modes, as shown Figure \ref{fig:prop}. The figure actually shows that the propagating mode significantly more sensitive to nonlocality because of its large wavevector $k_1$. A high wavevector actually means a large value for $\Omega$ and thus a noticeable effect on the dispersion relation.

\begin{figure}[htb]
\centerline{\includegraphics[width=8cm]{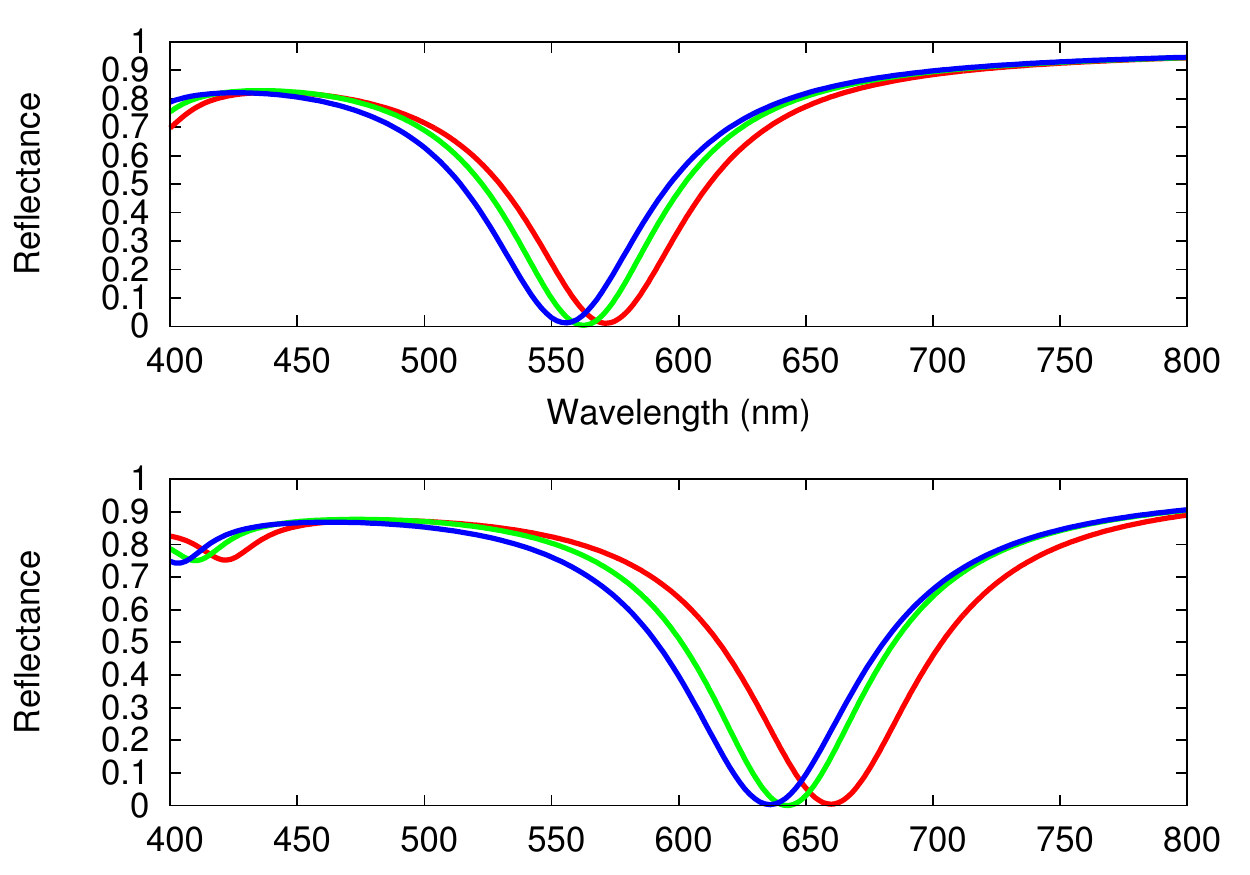}}
\caption{(Color Online) Reflection coefficient computed using COMSOL (fully nonlocal calculation, blue curves), using a local 2-modes model (red curves) and using a nonlocal version of the 2-modes model (green curves) for 3 nm wide grooves (top) and 2 nm wide grooves (bottom).\label{fig:final}}
\end{figure}

In order to assess the impact of nonlocality on the whole structure, we have used COMSOL simulations, as a full nonlocal modal method is still beyond our reach. As expected, the resonance is blue-shifted compared to a local calculation (from 15 nm for 3 nm wide slits, see \ref{fig:final} (b) to 24 nm for 2 nm wide slits, see Fig. \ref{fig:final} (a)). Interestingly, using a two-mode model but {\em changing only the propagation constant $k_1$ of the fundamental mode} as computed using the nonlocal dispersion relation \eqref{eq:dispnl} allows to account for most of the shift (see Fig. \ref{fig:final}). This demonstrates that the major reason why the whole structure is sensitive to the spatially dispersive response of the metal is that nonlocality has an impact on the wavevector of the single mode propagating in the slits. Its wavevector is actually not as high as predicted by the local theory, thus leading to a blue shift of the resonance.

\section{Conclusion}

The mode propagating in a single extremely thin slit is often called a gap-plasmon polariton, to distinguish it from the surface plasmon polariton. The propagating mode considered here is of course closely related and is submitted to the same physical effect: the presence of the metal makes the mode slow, with a very large wavevector. This allows to understand (i) that deeply subwavelength structures, sometimes smaller than the skin depth, can still be considered as cavities and (ii) why, as we have shown here, the smallest resonators are likely to be influenced by spatial dispersion in metals. This class of resonators are called gap-plasmon resonators and it has recently been demonstrated experimentally that these resonators could constitute extraordinarily efficient concentrators and absorbers\cite{polyakov11,polyakov12,moreau12b,siegfried13,blackgold} and produce totally unprecedented Purcell enhancements\cite{akselrod14,hoang15}. Theoretical studies show that they have a lot of potential as scatterers for light extraction too\cite{jung09,jouanin14}. We are thus confident that the very general theoretical tools we have introduced here can be useful to study these structures and help design them in the future, as they reach sizes that are well below the size of conventional cavity resonators.

\section*{Acknowledgments}

This work has been supported by the Agence Nationale de la Recherche (ANR), project ANR-13-JS10-0003.

\bibliography{nonlocal}

\end{document}